# Software Cognitive Information Measure based on Relation Between Structures

Yong-Hwa Choe, Chol-Yong Jong and Song Han

Faculty of mathematics, **Kim Il Sung** university, Pyongyang, D.P.R.Korea

**Abstract:** Cognitive complexity measures quantify human difficulty in understanding the source code based on cognitive informatics foundation. The discipline derives cognitive complexity on a basis of fundamental software factors i.e, inputs, outputs, and internal processing architecture. An approach to integrating Granular Computing into the new measure called Structured Cognitive Information Measure or SCIM. The proposed measure unifies and re-organizes complexity factors analogous to human cognitive process. However, according to the methodology of software and the scope of the variables, Information Complexity Number(ICN) of variables is depended on change of variable value and cognitive complexity is measured in several ways. In this paper, we define the Scope Information Complexity Number (SICN) and present the cognitive complexity based on functional decomposition of software, including theoretical validation through nine Weyuker's properties.

**Keywords**: Cognitive complexity measure, scope information complexity, scope information, Basic Control Structure(BCS),granule,BCS unit

**2010 Mathematics Subject Classification**: 68Q25, 68W40

## 1. INTRODUCTION

COGNITIVE complexity measures attempt to quantify the effort or degree of difficulty in comprehending the software based on cognitive informatics foundation that "cognitive complexity of software is dependent on three fundamental factors: inputs, outputs, and internal processing" [10]. Since then, many approaches [1, 3, 4, 5, 9, 10, 11] have been modified from 'Cognitive Function Size '(CFS) to fully consider complexity factors, e.g. include the evaluation of information contained in software as suggested by the informatics laws of software that "complexity of software is in the form of difficulty in understanding the information contained".

In [2], they observed that a variable accumulates the complexity from its preceding occurrences where it was assigned the value, as its value depends on those preceding appearances. Since they had to focus on particular granule when evaluating its complexity, we include the complexity from the variable's occurrences in preceding granules into the variable itself, so that they focused on its current occurrence in the granule that are being evaluating. This paper therefore aims to propose an approach to scope information complexity number of variable and scope information of program by applying

BCS unit computing strategies, which has been recently suggested as a problem-solving paradigm analogous to human cognitive process, so that the complexity metric can be derived more rigorously. We define the Scope Information Complexity Number (SICN) and present the cognitive complexity based on functional decomposition of software, including theoretical validation through nine Weyuker's properties. The contents in this paper are organized as follows: section 2 reviews existing cognitive complexity measures. Then we define the Scope Information Complexity Number (SICN) and Scope Information of program in section 3, Then we relate BCS unit computing strategies to cognitive complexity measurement to present our modified measure in section 4, followed by theoretically validating our measure through Weyuker's properties in section 5.

## 2. EXISTING COGNITIVE COMPLEXITY MEASURES

- *Cognitive Functional Size* (CFS)

Wang[10] defined the *cognitive functional size* (CFS) as follows:

$$CFS = (N_i + N_o) * W_c.$$

Here $N_i$ is the number of input variables and $N_o$ is the number of output variables. $W_c$ is the total cognitive weight of *basic control structures*(BCSs) and it is defined as the total sum of cognitive weights of its *q* linear blocks composed of individual BCSs. Since each linear block consists of *m* layers of nesting BCSs and each layer consists of *n* linear BCSs, then

$$W_c = \sum_{j=1}^{q} [\prod_{k=1}^{m} \sum_{i=1}^{n} W_c(j,k,i)]$$

Table 1

| Category | BCS | $W_c$ |
|---|---|---|
| Sequence | SEQ | 1 |
| Branch | If-Then-Else(ITE) | 2 |
|  | Case(CASE) | 3 |
| Iteration | For-do | 3 |
|  | Repeat-until | 3 |
|  | While-do | 3 |
| Embedded Component | Function call | 2 |
|  | Recursion | 3 |
| Concurrency | Parallel | 4 |
|  | Interrupt | 4 |

CFS interestingly triggered the study in software complexity measurement based on cognitive informatics foundation that complexity of software is dependent on inputs, outputs, and its internal processing. However, the factors were not "fully considered" [3, 4, 5], leading to many attempts to take into account the information content in the form of identifiers, operators, or variable references.

Kushwaha and A.K. Misra [3] modified Wang's CFS to measure the information contained in software. They referred to the law of informatics that software ≈ information, thus difficulty in understanding software ≈ difficulty in understanding information, and since software is a mathematical entity that represents computational information, the amount of information contained in software is a function of identifiers that hold the information and operators that perform the operations on the information. Hence:

$$Information = f(Indentifiers, Operators)$$

- *Cognitive Information Complexity Measure*

The Cognitive Information Complexity Measure (CICM) was then defined as:

$$CICM = WICS * W_c$$

where $W_c$ is the same as in CFS, and WICS is the weighted information count of the software derived from:

$$WICS = \sum_{k=1}^{LOCS} \{n(k)/(LOCS-k)\}$$

where n(k) is the number of identifiers and operators in the $K^{th}$ line.

- *Modified Cognitive Complexity Measure* (MCCM)

Later in 2006, S. Misra [4] also modified CFS into Modified Cognitive Complexity Measure, simplifying the complicated weighted information count in CICM:

$$MCCM = (N_{i1} + N_{i2}) * W_c$$

where $N_{i1}$ is the total number of occurrences of operators, $N_{i2}$ is the total number of occurrences of operands, and $W_c$ is the same as in CFS.

However, the multiplication of information content with the weight $W_c$ derived from the *whole* BCS's structure remains the approach's drawback.

In 2007, S. Misra proposed CPCM [4] based on the arguments that the occurrences of inputs/outputs in the program directly affect the internal architecture and are the forms of information contents. He also criticized the computation of CFS that the multiplication of distinct number of inputs and outputs with the total cognitive weights was not justified as there was no reason why using multiplication.

Furthermore, he claimed that operators are run time attributes and cannot be regarded as information contained in the software as proposed by CICM. Based on these arguments,

CPCM was thus defined as

$$CPCM = S_{io} + W_c$$

where $S_{io}$ is the total occurrences of input and output variables and $W_c$ is as in CFS.

In [2], they observed that a variable accumulates the complexity from its preceding occurrences where it was assigned the value, as its value depends on those preceding appearances. Since they had to focus on particular granule when evaluating its complexity, we include the complexity from the variable's occurrences in preceding granules into the variable itself, so that they focused on its current occurrence in the granule that are being evaluating. Their strategies followed by:

*At the beginning of the program, the Informatics Complexity Number (ICN) of every variable is zero. When a variable is assigned the value in the program, its ICN increases by 1, and if that assignment statement contains operators, ICN of the variable that is assigned the value also further increases by the number of operators in that statement.*

*For variable 'V', L is a program, $ICN_{max}(V,L)$ is the highest ICN of V 's occurrences in L. I(L) is defined as the sum of $ICN_{max}(V,L)$ of every variable V exists in L.*

**Eg1.**
```
public static void main(string[] args)
{ int UserInput;
   int square;
      UserInput=Text1.getInt();
      Square=UserInput*UserInput;
      System.out.print(square);
}
```
In Eg.1,
   ICN(UserInput)=1
   ICN(square)=2
   I(L)=1+2=3

Cognition of variables based on the scope of variables. When a variable is used as private variable and public variable, the variable is used as different function in different action scope, therefore its meaning of information is not same.

**Eg 2.**
```
int amount=123;//public variable
amount=amount*2;
void main()
{
    int amount=456;//local variable
    amount=amount+1;
    cout<<::amount;// public variable
    {
        int amount=789;//other local variable
        amount=amount--;
        cout<<::amount;// public variable that is out of method main
           //variable is not 456
        cout<<amount;//amount is local variable 789
    }
```

}

In Eg 2, amount variable is public variable as well as local variable.

According to variable's scope, its meaning is different and it must be comprehended each case.

In [2], they decompound software with BCS units and calculated software complexity by multiplying cognitive weight and information of BCS units based on information complexity number of variables.

Information of BCS unit is amount of value change of variable in granule (BCS unit), therefore we must consider value change of variable in granule.

However, [2] considered value change of variable in whole scope from beginning to granule (BCS unit) considered. This way cannot express granule information.

We extend Information Complexity Number (ICN) to Scope Information Complexity Number (SICN) and present the cognitive complexity based on functional decomposition of software.

And we decompound software with BCS units and define functional cognitive complexity based on Information and cognitive weight of granules-BCS units

## 3. Scope Information Complexity Number of Variables

In order to measure more cognitive and comprehensive complexity with the scope of variables and decomposition methodology of BCS unit of software, we define the concept of scope of variables and scope information of program followed by:

[**Definition 1**] Scope Information Cognitive Number of Variables ( $SICN$ )

① When each variable of software is introduced, its $SICN$ is 1 at the beginning of its action scope.

② When a variable is assigned value, its $SICN$ increases by 1.

If that assignment statement contains operators, $SICN$ of the variable that is assigned the value also further increases by the number of operators in that statement.

③ When scope of a variable is changed, perform of ② stops and save $SICN$ of the variable.

For new action scope, the variable is recognized to another new variable and we calculate its $SICN$ in the new action scope according to ①,②

④ Ending the new action scope, for $SICN$ saved of the variable we continue the perform operation of ②.

*L is program* or *part of program.*

[**Definition 2**] *For variable 'V' appearing in L, $SICN_{max}(V,L)$ is the highest SICN of V's occurrences in L. For variable 'V' appearing in L, $SICN_{min}(V,L)$ is the lowest SICN of V's occurrences in L.*

[**Definition 3**] *Scope Information contained in L (SI(L))*

*Scope Information contained in L (SI(L)) is defined as the sum of $[SICN_{max}(V,L) - SICN_{min}(V,L)]$ of every variable V exists in L.*

$$SI(L) = \sum_{V \in L}(SICN_{max}(V,L) - SICN_{min}(V,L))$$

*Scope Information* is defined value change number of variable in some scope to express cognition of variable.

Cognition Information of program '*L*' is value change number of variable in *L*.

**Eg 3.**
```
#include<ostream.h>
void main()
{    int key[]={20,10,50,40,60,70,30,45,67,15};
     int n=10,s=0;
     int left=1,right=n,m=n,buf;
     for(int i=0;i<n;i++)
     s=s+key[i];
     while(left<right)
     { int i=n;
         while (i>0)
         {
             if(key[i-1]>key[i])
             {   buf=key[i-1];
                 key[i-1]=key[i];
                 key[i]=buf;
             }                                              L1
             m=i;
             i=i-1;
         }
         left=m+1;
         for(int s=left;s<=right;s++)
         {
             if(key[s-1]>key[s])
             {   i=i+key[s];
                 buf=key[s-1];
                 key[s-1]=key[s];
                 key[s]=buf;
                 s=s-1;                                     L2
             }
             m=s;
         }
         right=m-1;
     }
     for(int i=0;i<n;i++)
     cout<<key[i]<<"\t";
     cout<<::s;
}
```

In Eg 3,this shows comparison of $ICN$ and $SICN$.

$$ICN_{\max}(s,L1)=3, \quad SICN_{\max}(s,L1)=3, \quad ICN_{\max}(s,L2)=8, \quad SICN_{\max}(s,L2)=5$$
$$W_{BCS}(L1)=4, \quad W_{BCS}(L2)=4$$

∘ complexity by $ICN$
$$W_{BCS}(L1)*I(L1)+W_{BCS}(L2)*I(L2)=44$$

∘ complexity by $SICN$
$$W_{BCS}(L1)*SI(L1)+W_{BCS}(L2)*SI(L2)=32$$

## 4. Cognitive Information Complexity of Software : ESCIM
### (Extended Structural Cognitive Information Measure)

*A. Decomposition of Software into BCS Hierarchical Structure*

In this section, we suggest extended Structural Cognitive Information Measure based on scope information complexity of variables and BCS unit decomposition of software.

To apply granular computing strategies to cognitive complexity measurement, first we decompose software into a hierarchy of granules.

When we comprehend the software, a BCS can be seen as a comprehension unit of which we need to understand functionalities and inputs/outputs before understanding interaction between BCS units and the whole program. Therefore, in the context of cognitive complexity measurement, we view a granule as a basic control structure (BCS), which may contain nested inner BCS's and information content.

The decomposition methodology of the program can be explained as followed:

  1) At the top level of the hierarchy, the whole program is partitioned into granules of BCS's in linear structure.

  2) Each granule whose corresponding BCS contains nested BCS's inside, is further partitioned generating next level of hierarchy.

  3) The partitioning stops when corresponding BCS to the granule is a linear BCS.

In brief, each level of the hierarchy consists of BCS's in linear structure, and because a BCS that contains no nested BCS's inside can be said to contain a single linear BCS, leaf nodes of the decomposed hierarchy are the linear BCS's. An example construction of the hierarchy from a program from [2] can be demonstrated as in Fig 1.

Eg 4.

```
public static void main(string[] args)
{                                                    G1
    int []numbers;
    int numcount;
    int num;
    numbers=new int[100];
    numcount=0;
    Text1.putln("Enter 10 integers");
```

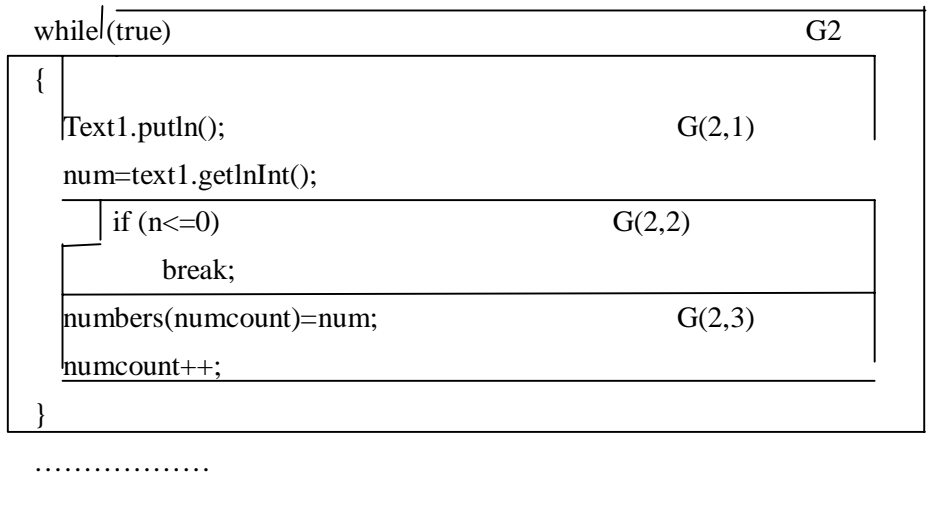

```
        while (true)                                      G2
        {
            Text1.putln();                                G(2,1)
            num=text1.getlnInt();
               if (n<=0)                                  G(2,2)
                    break;
            numbers(numcount)=num;                        G(2,3)
            numcount++;
        }
        ………………
   }
```

Demonstration of BCS hierarchical structure construction of Eg 4 is followed by:

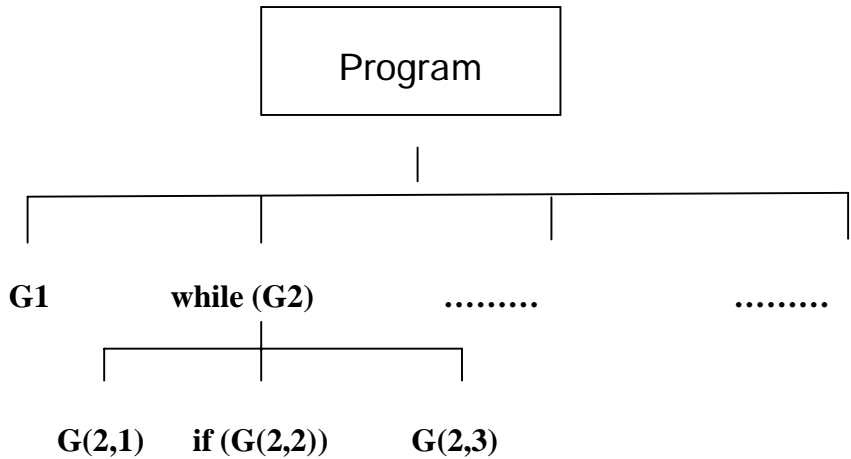

B. *The Extended Structural Cognitive Information Measure of Software* (ESCIM)

Definition 4. ESCIM is defined as the total sum of the products of corresponding cognitive weights and information contained in leaf node granule (I(L)). Since software may consist of q linear blocks composed in individual BCS 's, and each block may consist of 'm ' layers of nesting BCS's, and each layer with 'n ' linear BCS 's, then q m n

$$ESCIM = \sum_{j=1}^{q} W_c(j) [\prod_{k=1}^{m} \sum_{i=1}^{n} SI(j,k,i) * W_c(j,k,i)]$$

where weights $W_c(j,k)$ of BCS's are cognitive weights of BCS's presented in [2], and $SI(j,k,i)$ are information contained in a leaf BCS granule as defined in Definition 3.

From the definition, we can say that ESCIM evaluates the complexity by taking into account the dependencies of variables and their position in the BCS's structure as suggested by Fig 1.. Number of inputs/outputs can now be disregarded as 1I0s variables have already been included as the information contained in the program.

C. *The Unit of ESCIM*

In ESCIM, the simplest software component with only one variable assignment, no operators, and a linear sequential BCS structure, is defined as the Extended Structural Cognitive Information unit (ESCIU), computing ESCIM can be formulated as:

$$ESCIM = 1 * 1 = 1 \ [ESCIU]$$

The value in SSCU of a software system indicates its cognitive complexity relative to that of the defined simplest software component,

$$ESCIU = \frac{\text{cognitive complexity of the system}}{\text{cognitive complexity of the defined simplest software component}}$$

## 4. VALIDATION THROUGH WEYUKER PROPERTIES

The proposed ESCIM can be proved to satisfy all nine Weyuker's properties, which are often used to evaluate and compare complexity measures as shown in Table 2.

COMPARISON OF CONFORMANCE OF COMPLEXITY MEASURES TO WEYUKER'S PROPERTIES

**Table 2**

| Property | LOC | McCabe's Cyclomatic | Halstead's Effort | Dataflow Complex | CFS | MCCM | CPCM | SCIM | ESICM |
|---|---|---|---|---|---|---|---|---|---|
| 1 | / | / | / | / | / | / | / | / | / |
| 2 | / |   | / | × | / | / | / | / | / |
| 3 | / | / | / | / | / | / | / | / | / |
| 4 | / | / | / | / | / | / | / | / | / |
| 5 | / | / | × | × | / | / | / | / | / |
| 6 | × | × | / | / | × | × | × | / | / |
| 7 | × | × | × | / | / | × | × | / | / |
| 8 | / | / | / | / | / | / | / | / | / |
| 9 | × | × | / | / | / | / | / | / | / |

Let P and Q be program body.

**Property 1**. (∃**P**)(∃**Q**)(|**P**|≠|**Q**|)

This property states that the measures should not rank all the programs as equally complex. Therefore, ESCIM obviously satisfies this property.

**Property 2**. (∀**P**),|**P**|≥0

Let c be a nonnegative number, then there are only finitely many programs of complexity c.

Since all programming languages can have only finite number of BCS's, variable assignments, and operators, it is assumed that some largest numbers can be used as an upper bound on the numbers of BCS's, variable assignments and operators. Therefore, for these numbers, there are finite many programs having that much number of BCS's, variable assignment, and operators. Consequently, for any given value of ESCIU, there exists finitely large number of programs, and ESCIM satisfies this property.

**Property 3**. (∃**P**)(∃**Q**)(|**P**|=|**Q**|)

There are distinct program P and Q such that !PI =IQI.

ESCIM clearly satisfies this property as at least for any program containing operator '+', replacing '+' with '-' will result in a different program with the same ESCIM complexity.

**Property 4**. (∃**P**)(∃**Q**)(P=Q & |**P**|≠|**Q**|)

This property states that there exist two programs equivalent to each other (i.e. for all inputs given to the program, they halt on the same values of outputs.) with different complexity.

Clearly, the program computing 1+2+…+n can be implemented with while loop, or simply sequence structure with formula n(n+1)/2. The values of ESCIM from these two implementations are different. Hence, ESCIM satisfies this property.

**Property 5**. (∃**P**)(∃**Q**)(|**P**|≤|**P;Q**| & |**Q**|≤|**P;Q**|)

ESCIM obviously satisfies the property because adding any program body whether to the end or before the beginning of a program body can only increase or hold the ESCIM complexity.

**Property 6a**. (∃**P**)(∃**Q**) (∃**R**) (|**P**|=|**Q**| & |**P;R**|≠|**Q;R**|)

Given program P and Q with same value in ESCIU, and program R contains some variables that are assigned values in P but no variables that are assigned values in Q, IP;RI is clearly more than IQ;RI because SICNs of those variables in R of P;R are higher than those of the same variables in R of Q;R. Therefore, ESCIM satisfies this property.

**Property 6b**. (∃**P**)(∃**Q**) (∃**R**) (|**P**|=|**Q**| & |**R;P**|≠|**R;Q**|)

In the same way as in property 6a, ESCIM satisfies this property.

The satisfaction of property 6 indicates one strength of ESCIM over other cognitive complexity measures that when different programs with the same complexity value are extended with the same program part, other measures view the extended programs as having the same complexity no matter what. This is because they do not consider possible complexity transferred between BCS in linear structures, or view linear BCS's as completely separately comprehensible, while ESCIM estimates the complexity transferred between blocks of BCS by the cumulative variable complexity counting scheme and does not overlook interrelationships among granules.

The intent behind Weyuker's Properties is to check whether complexity value of a program is suitable with complexity values of its parts. However the definitions leave some room for measures to slip through. For example, CICM happens to satisfy Property 6 because its weighing of information content is so random that there exist programs P, Q, R that IPI=IQI but IP;RI ≠ IQ;RI. Even though sometimes, if R is completely independent of P and Q, IP;RI should be the same as IQ;RI. We can say that the measure that truly satisfies the intent of Weyuker's properties should be able to answer what would happen to IP;RI when P and R are in some condition to each other. For ESCIM, IP;RI equals to IPI+IRI when cognition of R in IP;RI is not affected by P, while IP;RI > IPI+IRI when P has some effects on R.

**Property 7**. There are some program bodies P and Q such that Q is formed by permuting the order of statements of P, and |**P**|≠|**Q**|.

ESCIM satisfies this property because the permutation of statements can result in different SICNs, hence making the ESCIM value different.

**Property 8**. If P is renaming of Q, then IPI = IQI

ESCIM clearly satisfies this property as it does not take into account the names.

**Property 9**. (∃**P**)(∃**Q**)(|**P**|+|**Q**|≤|**P; Q**|)

ESCIM satisfies this property because if some variables assigned values in P occur in Q, the complexity of Q in P;Q will increase from Q alone because the SICNs of those variable will increase, hence making IP;QI higher than IPI + IQI.

- CFS, SCIM and ESCIM can indicate the coding efficiency (E), which can be defined as:

$$E = \frac{ESCIM}{LOC}$$

The higher coding efficiency indicates the higher complexity information packed in the shorter program code, therefore the program is likely to contain more defects than the program with lower coding efficiency.